\documentclass[lettersize, journal]{IEEEtran}
\usepackage{amssymb}
\usepackage{amsmath}
\usepackage{cite}
\usepackage{url}
\usepackage{xcolor}
\usepackage{cite,graphicx,amsmath,amssymb}
\usepackage{subfigure}
\usepackage{citesort}
\usepackage{fancyhdr}
\usepackage{mdwmath}
\usepackage{mdwtab}
\usepackage{caption}
\usepackage{amsthm}
\usepackage{setspace}
\usepackage{algorithm}
\usepackage{algorithmic}
\usepackage{makecell}
\usepackage{diagbox}
\usepackage{amsmath}

\newtheorem{theorem}{Theorem}

\newtheorem{lemma}{Lemma}

\newtheorem{corollary}{Corollary}

\allowdisplaybreaks
\makeatletter
\def\ScaleIfNeeded{
\ifdim\Gin@nat@width>\linewidth \linewidth \else \Gin@nat@width
\fi } \makeatother
 
\begin{document}
\title{SE and EE Tradeoff in Active STAR-RIS Assisted Systems With Hardware Impairments}
\author{Ao~Huang, Xidong~Mu, Li~Guo, and Guangyu~Zhu
\thanks{Ao Huang, Li Guo and Guangyu Zhu are with the School of Artificial Intelligence, Beijing University of Posts and Telecommunications (BUPT), Beijing 100876, China, also with the Key Laboratory of Universal Wireless Communications, Ministry of Education, BUPT, also with the Engineering Research Center of Blockchain and Network Convergence Technology, Ministry of Education, BUPT, also with the National Engineering Research Center for Mobile Internet Security Technology, BUPT, and also with the Hainan Advanced Digital Technology and System Laboratory, BUPT (email: huangao@bupt.edu.cn; guoli@bupt.edu.cn; zhugy@bupt.edu.cn).}
\thanks{Xidong Mu is with the Centre for Wireless Innovation (CWI), Queen's University Belfast, Belfast, BT3 9DT, U.K. (e-mail: x.mu@qub.ac.uk).}
}

\maketitle

\begin{abstract}
This paper investigates the problem of resource efficiency maximization in an active simultaneously transmitting and reflecting reconfigurable intelligent surface (STAR-RIS) assisted communication system under practical transceiver hardware impairments (HWIs). We aim to obtain an optimal tradeoff between system spectral efficiency (SE) and energy efficiency (EE), by jointly optimizing the base station (BS) transmit beamforming and the active STAR-RIS beamforming. To tackle the challenges in the fractional objective function, we begin by applying the quadratic transformation method to simplify it into a manageable form. An alternating optimization-based algorithm is then developed to iteratively update the BS and STAR-RIS beamforming coefficients. Simulation results demonstrate that the proposed scheme performs better than other baseline schemes in the presence of HWIs. Moreover, the variation of the achievable SE-EE region with different transmit power budgets is analyzed.
\end{abstract}
\vspace{-0.1in}
\section{Introduction}
Of late, simultaneously transmitting and reflecting reconfigurable intelligent surfaces (STAR-RISs) have attracted considerable interest as a revolutionary technology for next-generation wireless networks, due to their capacity to dynamically shape wireless channel conditions~\cite{mu2021simultaneously}. By properly adjusting the phase shift of each element on the STAR-RIS, users on both sides of the surface can experience enhanced performance. Driven by the unique advantages of STAR-RISs, a growing body of research has been steadily dedicated to STAR-RIS assisted systems~\cite{liu2021star,xu2022simultaneously,mu2024simultaneously}.

It is important to point out that most existing studies assume ideal hardware at the transceiver. Nevertheless, in practical communication systems, both transmitters and receivers are inevitably affected by non-ideal characteristics, such as I/Q imbalance, amplifier non-linearities, and quantization errors~\cite{bjornson2014massive}. Despite this, only a few initial works have begun to explore the impact of hardware impairments (HWIs) in STAR-RIS assisted systems~\cite{li2023achievable,li2024ergodic,zhou2023active}. Specifically, the authors of~\cite{li2023achievable} considered a fundamental two-user setup assisted by a passive STAR-RIS, taking into account transceiver HWIs and focusing on maximizing the system achievable rate with imperfect channel state information (CSI). Building on this, in~\cite{li2024ergodic}, the system model was extended to incorporate a multi-antenna base station (BS). With the aim of improving the system ergodic rate, a promising two-timescale protocol for joint beamforming design was introduced. Capitalizing on the potential of active STAR-RISs, the authors of~\cite{zhou2023active} suggested deploying an active STAR-RIS to assist transmission in a symbiotic radio communication system, where wireless resource allocation was optimized for minimizing the total power consumption.

Active STAR-RISs, with their signal amplification capabilities, can better compensate for performance degradation caused by transceiver HWIs.
However, it is often overlooked that the non-linear correlation between system spectral efficiency (SE) and energy efficiency (EE) presents a great challenge in optimizing both metrics simultaneously. Although active STAR-RISs can well alleviate the ``multiplicative fading'' effect of signals~\cite{zhang2022active}, the amplification power required by the active elements inevitably causes a rise in overall system energy consumption. Meanwhile, the impact of HWIs is also amplified, leading to more significant signal distortions at both the transmitter and receivers. To mitigate these impairments and achieve satisfactory data rates without substantial power consumption increases, more advanced system design strategies are imperative. Our motivation lies in addressing the complex tradeoff between SE and EE, which becomes particularly prominent in active STAR-RIS assisted multiuser systems, especially when considering transceiver HWIs and the amplified noise induced by the active STAR-RIS. To our knowledge, no prior work has fully explored this area, which motivates our research.

In light of the above, we consider an active STAR-RIS assisted communication system where HWIs are present at both the transmitter and receivers. To achieve an optimal system SE-EE tradeoff, we formulate a resource efficiency (RE) maximization problem by jointly designing the transmit beamforming at the BS and the configuration of each active STAR-RIS element. In particular, the quadratic transformation method is applied to reformulate the fractional objective into a more tractable form. Furthermore, we develop an alternating optimization (AO)-based algorithm to iteratively optimize the BS precoding and STAR-RIS configuration. Finally, simulation results validate the viability of the proposed system model and optimization framework.
\section{System Model and Problem Formulation}
\subsection{System Model}
\begin{figure}[t]
	\centering
	\includegraphics[width=3.2in]{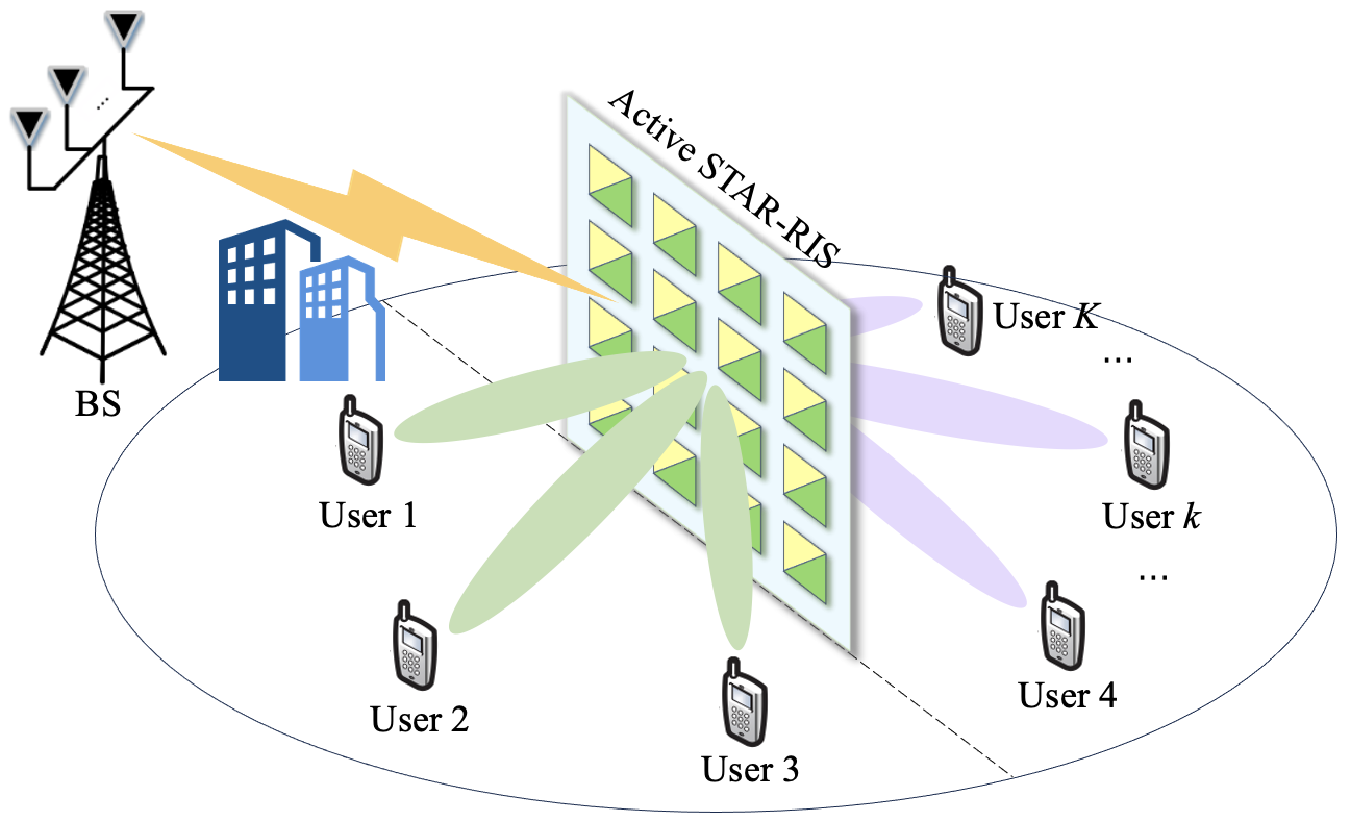}
	\caption{Illustration of the active STAR-RIS assisted multiuser communication system with transceiver HWIs. }
	\label{}
\end{figure}
As illustrated in Fig. 1, we consider a downlink communication system assisted by an active STAR-RIS, with transceiver HWIs taken into account. In this setup, an active STAR-RIS facilitates the communication between a BS equipped with $N$ antennas and multiple single-antenna users. There are $K$ users, which are denoted by the set $\mathcal{K}=\{1,2,\cdots,K\}$. Depending on where they are, the users are split into two groups. Let the set $\mathcal{R}$ represent the users located in the reflection region of the STAR-RIS, while the set $\mathcal{T}$ includes those located in the transmission region. The active STAR-RIS is composed of $M$ elements, and $\mathcal{M}=\{1,2,\cdots,M\}$ is the set comprising all of the STAR-RIS elements.
Each element is assumed to operate in the energy splitting (ES) mode~\cite{mu2021simultaneously}. Let $\mathbf{\Theta}_t={\textup{diag}}\left(\mathbf{u}_t^H\right)\in\mathbb{C}^{M \times M}$ and $\mathbf{\Theta}_r={\textup{diag}}\left(\mathbf{u}_r^H\right)\in\mathbb{C}^{M \times M}$ denote the diagonal transmisison and reflection coefficient matrices of the STAR-RIS with $\mathbf{u}_s=[u^s_1, u^s_2,\cdots,u^s_M]^H\in\mathbb{C}^{M \times 1}$ and $u^s_m=\sqrt{\beta_m^s}e^{j\theta_{m}^s}$, $\forall s\in\{r,t\}$, where $\sqrt{\beta_m^s}$ and $\theta_{m}^s, \forall m\in\mathcal{M}$ denote the amplitude and phase-shift coefficients of the $m$-th element, respectively. Based on the amplitude and phase-shift characteristics of each element, the feasible set of $u^s_m$ can be defined as$\footnote{We note that the coupled phase-shift condition is primarily required for passive and lossless STAR-RISs~\cite{wang2022coupled}. In contrast, our work focuses on an active STAR-RIS, which provides greater flexibility in adjusting transmission and reflection phase shifts of each element independently.}$
\begin{equation}
	{\Phi}=\left\{u^s_m| \beta_m^s\le \rho_{\max}, \theta_{m}^s\in [0,2\pi)\right\},
\end{equation}
where $\rho_{\max}>1$ denotes the maximum power gain achievable by the active load.

We assume that the direct communication links between the BS and the users are blocked by obstacles. For the purpose of exposition, let $\mathbf{G}\in\mathbb{C}^{M \times N}$ and $\mathbf{h}_{k}^H\in\mathbb{C}^{1\times M}$ denote the channels from the BS to the active STAR-RIS and the STAR-RIS to user $k \in\mathcal{K}$, respectively. The analysis in this work is conducted with the assumption of perfect CSI to investigate the impact of transceiver HWIs on system performance~\cite{wu2021channel}.
\vspace{-0.4cm}
\subsection{Signal Model}
The transmitted signal from the BS is given by
\begin{equation}\mathbf{x}=\sum_{k\in\mathcal{K}}\mathbf{w}_ks_k+\mathbf{z}_b,
\end{equation}
where $s_k$ denotes the information-bearing signal intended for user $k$ with $\mathbb{E}\{|s_k|^2\}=1$ and $\mathbf{w}_k\in\mathbb{C}^{N \times 1}$ denotes the precoding vector. $\mathbf{z}_b\sim \mathcal{CN}\left(\mathbf{0},\kappa_b\sum_{k\in\mathcal{K}}{\rm{Tr}}\left(\mathbf{w}_k\mathbf{w}_k^H\right)\right)$ is the transmitter distortion noise, which models the aggregate residual HWIs at the BS. Here, $\kappa_b\in[0,1)$ is the proportionality coefficient which characterizes the level of the HWIs.

The signal received at user $k\in\mathcal{K}$ is represented by
\begin{equation}
	\begin{aligned}
		y_{k}&=(\mathbf{h}_{k}^H\mathbf \Theta_{s_k}\mathbf{G})\mathbf{x}+\mathbf{h}_{k}^H\mathbf\Theta_{s_k}\mathbf{n}_{A}+n_k+n_{u,k}\\
		&\triangleq \tilde{y}_{k}+n_{u,k}, \forall k\in\mathcal{K},
	\end{aligned}
\end{equation}
where $s_k\in\{t,r\}$ represents the half-space of the STAR-RIS where user $k$ is located. Specifically, $s_k=t$ if $k\in\mathcal{T}$, and $s_k=r$ otherwise. $\mathbf{n}_{A}\sim \mathcal{CN} (\mathbf{0},
\sigma_{A}^2\mathbf{I}_M)$ is the thermal noise generated at the active STAR-RIS. $n_k\sim\mathcal{CN} (0, \sigma^2)$ is the additive white Gaussian noise (AWGN) at user $k$ with noise power $\sigma^2$. 
The HWIs at user $k$ are denoted by $n_{u,k}\sim\mathcal{CN} (0,
\kappa_u\mathbb{E}\{|\tilde{y}_{k}|^2\})$, where $\kappa_u\in[0,1)$ is the proportionality coefficient characterizing the level of the HWIs at users.

Consequently, the received signal-to-interference-plus-noise ratio (SINR) at user $k\in\mathcal{K}$ is expressed as
\begin{equation}
	\textup{SINR}_{k}=\frac{\left|\left(\mathbf{h}_{k}^H\mathbf \Theta_{s_k} \mathbf{G}\right)\mathbf{w}_k\right|^2}{\mathbb{E}\{\tilde{n}_{k}\tilde{n}_{k}^*\}}, \forall k\in\mathcal{K},
\end{equation}
where $\tilde{n}_{k}=\left(\mathbf{h}_{k}^H\mathbf \Theta_{s_k} \mathbf{G}\right)\left(\sum_{i\in\mathcal{K}\backslash\{k\}}\mathbf{w}_is_i+\mathbf{z}_b \right)+\mathbf{h}_{k}^H\mathbf \Theta_{s_k}\mathbf{n}_{A}+n_k+n_{u,k}$. Define the cascaded channel form the BS to user $k$ assisted by the active STAR-RIS as $\mathbf{H}_{k}^H\triangleq \mathbf{h}_{k}^H\mathbf \Theta_{s_k} \mathbf{G}$. Then, the power of the interference noise  i.e., ${\mathbb{E}\{\tilde{n}_{k}\tilde{n}_{k}^*\}}$, can be calculated by Eq. (5) at the top of the next page.
\begin{figure*}[h]
	\normalsize
	\begin{equation}
		\begin{aligned}
			\mathbb{E}\{\tilde{n}_{k}\tilde{n}_{k}^*\}=\sum_{i\in\mathcal{K}\backslash \{k\}}\left|\mathbf{H}_{k}^H\mathbf{w}_i\right|^2+\sum_{l\in\mathcal{K}}\mathbf{H}_{k}^H\left(\kappa_u\mathbf{w}_l\mathbf{w}_l^H+\left(1+\kappa_u\right)\kappa_b{\widetilde{\textup{diag}}}\left(\mathbf{w}_l\mathbf{w}_l^H\right) \right)\mathbf{H}_{k}+(1+\kappa_u)\left(\left\|\mathbf{h}_{k}^H\mathbf \Theta_{s_k}\right\|^2\sigma_{A}^2+\sigma^2\right).
		\end{aligned}
	\vspace{-0.3cm}
	\end{equation}
	\vspace*{-14pt}
\end{figure*}
The achievable data rate of decoding $s_k$ at user $k$ is obtained as
\begin{equation}
	R_{k}=\log_2\left(1+\textup{SINR}_{k}\right), \forall k\in\mathcal{K}.
\end{equation}
\vspace{-1cm}
\subsection{System SE and EE}
According to~\cite{mao2022rate}, the system SE is defined as the sum rate of all users, as given by Eq. (7) at the top of the next page.
\begin{figure*}[h]
	\small
	\begin{equation}
		\begin{aligned}
			\eta_{\rm SE}\triangleq\sum_{k\in\mathcal{K}}\log_2\left(1+\frac{\left|\mathbf{H}_{k}^H\mathbf{w}_k\right|^2}{\sum_{i\in\mathcal{K}\backslash \{k\}}\left|\mathbf{H}_{k}^H\mathbf{w}_i\right|^2+\sum_{l\in\mathcal{K}}\mathbf{H}_{k}^H\left(\kappa_u\mathbf{w}_l\mathbf{w}_l^H+\left(1+\kappa_u\right)\kappa_b{\widetilde{\textup{diag}}}\left(\mathbf{w}_l\mathbf{w}_l^H\right) \right)\mathbf{H}_{k}+(1+\kappa_u)\left(\left\|\mathbf{h}_{k}^H\mathbf \Theta_{s_k}\right\|^2\sigma_{A}^2+\sigma^2\right)}\right).
		\end{aligned}
	\vspace{-0.2cm}
	\end{equation}
	\hrulefill 
	\vspace*{-14pt}
\end{figure*}

With regard to the power consumption model, the transmit power at the BS can be written as
\begin{equation}
	P_{\rm BS}=\sum_{k\in\mathcal{K}}\left\|\mathbf{w}_k\right\|^2.
\end{equation}
Besides, the amplification power of the active STAR-RIS can be modelled as
\begin{equation}
	P_{act} =\sum_{s\in\{t,r\}}\left\|\mathbf\Theta_{s} \mathbf{G}\left(\sum_{k\in\mathcal{K}}\mathbf{w}_k+\mathbf{z}_b\right)\right\|^2
	+\sum_{s\in\{t,r\}}\sigma_{A}^{2}\left\|\mathbf{\Theta}_{s} \right\|^{2}_F.
\end{equation}
Therefore, the total power consumption of the system is calculated by
\begin{equation}
	P_{tot}= \frac{1}{\xi}\left(P_{\rm BS}+P_{act}\right)+MP_{r}+P_{c},
\end{equation}
where $\xi$ represents the efficiency of the power amplifier, $P_{r}$ accounts for the circuit power consumption at each STAR-RIS component, and $P_{c}$ indicates the additional power dissipation of the system. Then, the system EE is given by
\begin{equation}
	\eta_{\rm EE}\triangleq B\frac{\eta_{\rm SE}}{P_{tot}},
\end{equation}
where $B$ is the system bandwidth.
\vspace{-0.4cm}
\subsection{Problem Formulation}
Following~\cite{tang2014resource}, we introduce a unit-consistent performance metric, referred to as RE, to properly characterize the interplay between SE and EE, which is defined as follows:
\begin{equation}
	\eta_{\rm RE}\triangleq \frac{\eta_{\rm EE}}{B}+\omega\frac{\eta_{\rm SE}}{P_{\max}},
\end{equation}
where $\omega > 0$ is the weight representing the priorities of SE and EE. The denominator $P_{\max}=\frac{1}{\xi}\left(P_{\rm BS}^{\max}+P_{\rm RIS}^{\max}\right)+MP_{r}+P_{c}$ in the second term refers to the entire power budget of the considered system. Herein, $P_{\rm BS}^{\max}$ and $P_{\rm RIS}^{\max}$ denote the maximum transmit and amplification power at the BS and active STAR-RIS, respectively.

To solve the problem of maximizing the system RE, we propose to jointly optimize the transmit beamforming at the BS and the active beamforming at the STAR-RIS. Mathematically, the RE maximization problem is provided as
\begin{subequations}\label{problem formulation}
	\begin{align}	
		&\mathop{\rm{max}}\limits_{\mathbf{w}_{k},\mathbf{\Theta}_{s}}\quad \eta_{\rm RE}\left(\mathbf{w}_{k},\mathbf{\Theta}_{s}\right) \\
		&{\rm s.t.} 
		\sum_{k\in\mathcal{K}}\left\|\mathbf{w}_k\right\|^2\le P_{\rm BS}^{\max},\\
		& \quad\,
		\sum_{s\in\{t,r\}}\!\left\|\mathbf\Theta_{s} \mathbf{G}\!\left(\sum_{k\in\mathcal{K}}\mathbf{w}_k+\mathbf{z}_b\right)\!\right\|^2\!\!\!+\!\!\!\sum_{s\in\{t,r\}}\!\sigma_{A}^{2}\left\|\mathbf{\Theta}_{s} \right\|^{2}_F \!\le\! P_{\rm RIS}^{\max},\\
		& \quad\,
		R_{k}\ge R_{\min}, \forall k\in\mathcal{K},\\
		&\quad\,
		u^s_m\in\Phi, \forall s\in\{r,t\}, \forall m\in\mathcal{M},
	\end{align}
\end{subequations}
where (13b) and (13c) represent the corresponding power constraints. Constraint (13d) ensures that the achievable rate for each user does not fall below a specified rate threshold $R_{\rm min}$. (13e) defines the constraint associated with the configuration of the active STAR-RIS.
\vspace{-0.5cm}
\section{Proposed Solution}
\vspace{-0.2cm}
Generally, problem (13) is non-convex and nontrivial to solve in a straightforward manner. In this section, we first linearize the objective function in propblem (13) with the quadratic transformation method proposed in~\cite{shen2018fractional}. By introducing a slack variable $\alpha\in\mathbb{R}$, problem (13) can be equivalently recasted into a non-fractional formulation as:
\vspace{-0.2cm}
\begin{subequations}\label{problem formulation}
	\begin{align}	
		&\mathop{\rm{max}}\limits_{\mathbf{w}_{k},\mathbf{\Theta}_{s},\alpha} 
		2 \alpha \sqrt{\eta_{\rm SE}(\mathbf{w}_{k},\mathbf{\Theta}_{s})} - \alpha^{2} P_{tot}(\mathbf{w}_{k},\mathbf{\Theta}_{s})+\frac{\omega \eta_{\rm SE}(\mathbf{w}_{k},\mathbf{\Theta}_{s})}{P_{\max}}\\
		&\quad{\rm s.t.} \;\;\;\,\,
		\textup{(13b) -- (13e)}.
	\end{align}
\end{subequations}
The reformulation from (13) to (14) simplifies the original non-convex fractional problem by transforming it into a convex form that is more tractable. According to~\cite{shen2018fractional}, with given $\mathbf{w}_{k}$ and $\mathbf{\Theta}_{s}$, the optimal $\alpha$ can be readily derived as
\begin{equation}
	\alpha^{*}=\frac{\sqrt{\eta_{\rm SE}(\mathbf{w}_{k},\mathbf{\Theta}_{s})}}{P_{tot}(\mathbf{w}_{k},\mathbf{\Theta}_{s})}.
\end{equation}

However, problem (14) remains non-convex due to the strong coupling between the BS precoding vector $\mathbf{w}_{k}$ and the STAR-RIS beamforming matrix $\mathbf{\Theta}_{s}$. In the following part, we are going to iteratively optimize the variables $\mathbf{w}_{k}$ and $\mathbf{\Theta}_{s}$ with a given $\alpha$ based on the AO method.
\vspace{-0.8cm}
\subsection{Transmit Beamforming Design}
\vspace{-0.2cm}
For given $\mathbf{\Theta}_s$, we formulate the subproblem to optimize the transmit beamforming $\mathbf{w}_k$ at the BS as follows
\begin{subequations}\label{problem formulation}
	\begin{align}
		&\mathop{\rm{max}}\limits_{\mathbf{w}_{k}}\quad 
		2 \alpha \sqrt{\eta_{\rm SE}(\mathbf{w}_{k})} 
		-\alpha^{2} P_{tot}(\mathbf{w}_{k})+\frac{\omega \eta_{\rm SE}(\mathbf{w}_{k})}{P_{\max}}\\	
		&\;\;{\rm s.t.} \;\;\;\,\,
		\textup{(13b) -- (13d)}.
	\end{align}
\end{subequations}
We note that the constraints (13c) and (13d) are non-convex, which makes the transmit beamforming design particularly difficult.  Firstly, we can rewrite the transmit power required by the active STAR-RIS as
\begin{equation}
	\begin{aligned}
		P_{act} =&\sum_{s\in\{t,r\}}\left\|\mathbf\Theta_{s} \mathbf{G}\sum_{k\in\mathcal{K}}\mathbf{w}_k\right\|^2\\
		&+\sum_{s\in\{t,r\}}\kappa_b\textup{Tr}\left(\mathbf\Theta_{s} \mathbf{G}\left(\sum_{k\in\mathcal{K}}\widetilde{\textup{diag}}\left(\mathbf{w}_l\mathbf{w}_l^H\right)\right)\mathbf{G}^H\mathbf\Theta_{s}^H\right)\\
		&+\sum_{s\in\{t,r\}}\!\sigma_{A}^{2}\left\|\mathbf{\Theta}_{s} \right\|^{2}_F.
	\end{aligned}
\end{equation}
Given the fact that $\mathbf{B}^H\widetilde{\textup{diag}}(\mathbf{A} \mathbf{A}^H)\mathbf{B}=  \mathbf{A}^H\widetilde{\textup{diag}}(\mathbf{B} \mathbf{B}^H)\mathbf{A}$, the following constraint is obtained
\begin{equation}
	\sum_{k\in\mathcal{K}}\mathbf{w}_k^H\mathbf\Gamma\mathbf{w}_k+\sum_{s\in\{t,r\}}\sigma_{A}^{2}\left\|\mathbf{\Theta}_{s} \right\|^{2}_F \le P_{\rm RIS}^{\max},
\end{equation}
where $\mathbf\Gamma\triangleq\sum_{s\in\{t,r\}}\left(\mathbf{G}^H\mathbf\Theta_{s}^H\mathbf\Theta_{s}\mathbf{G}+\kappa_b\widetilde{\textup{diag}}\left(\mathbf{G}^H\mathbf\Theta_{s}^H\mathbf\Theta_{s}\mathbf{G}\right)\right)$. 

Define $\mathbf C_k\triangleq\kappa_u\mathbf{H}_k\mathbf{H}_k^H+\left(1+\kappa_u\right)\kappa_b\widetilde{\textup{diag}}\left(\mathbf{H}_k\mathbf{H}_k^H\right)$, the received SINR to decode its own signal at user $k$ can be equivalently expressed as Eq. (19).
\begin{figure*}[h]
	\normalsize
	\begin{equation}
		\begin{aligned}
	\textup{SINR}_{k}=\frac{\left|\mathbf{H}_{k}^H\mathbf{w}_k\right|^2}{\sum_{i\in\mathcal{K}\backslash \{k\}}\left|\mathbf{H}_{k}^H\mathbf{w}_i\right|^2+\sum_{l\in\mathcal{K}}\mathbf{w}_l^H\mathbf C_k\mathbf{w}_l+(1+\kappa_u)\left(\left\|\mathbf{h}_{k}^H\mathbf \Theta_{s_k}\right\|^2\sigma_{A}^2+\sigma^2\right)}.
		\end{aligned}
		\vspace{-0.3cm}
\end{equation}
\hrulefill 
\vspace*{-14pt}
\end{figure*}
Furthermore, we introduce slack variables $\{X_k\}$ and $\{Y_k\}$ such that
\begin{equation}
	\frac{1}{X_{k}}\le \left|\mathbf{H}_{k}^H\mathbf{w}_k\right|^2,
	\vspace{-0.1cm}
\end{equation}
\begin{equation}
	\begin{aligned}
	Y_{k}\ge &\sum_{i\in\mathcal{K}\backslash \{k\}}\left|\mathbf{H}_{k}^H\mathbf{w}_i\right|^2+\sum_{l\in\mathcal{K}}\mathbf{w}_l^H\mathbf C_k\mathbf{w}_l\\
	&\quad+(1+\kappa_u)\left(\left\|\mathbf{h}_{k}^H\mathbf \Theta_{s_k}\right\|^2\sigma_{A}^2+\sigma^2\right),
	\end{aligned}
\vspace{-0.3cm}
\end{equation}
for any given points $\{{X_{k}^{(t)}},{Y_{k}^{(t)}}\}$, based on the first-order Taylor expansion~\cite{boyd2004convex}, the concave lower bound of the achievable rate at user $k$ is acquired by
\begin{equation}
	\begin{aligned}
	R_{k}\ge&\log_2\left(1+\frac{1}{{X_{k}^{(t)}}{Y_{k}^{(t)}}}\right)-\frac{\left(\log_2e\right)\left({X_{k}}-{X_{k}^{(t)}}\right)}{{X_{k}^{(t)}}+{X_{k}^{(t)}}^2{Y_{k}^{(t)}}}\\
	&-\frac{\left(\log_2e\right)\left({Y_{k}}-{Y_{k}^{(t)}}\right)}{{Y_{k}^{(t)}}+{Y_{k}^{(t)}}^2{X_{k}^{(t)}}}\triangleq R_{k}^{lb}, \forall k \in\mathcal{K}.
	\end{aligned}
\end{equation}
However, there is one key observation that constraint (20) is non-covex and its right-hand side is a convex function with respect to $\mathbf{w}_k$. At the given local point $\mathbf{w}_k^{(t)}$, we have
\begin{equation}
	\begin{aligned}
\left|\mathbf{H}_{k}^H\mathbf{w}_k\right|^2&\ge\left|\mathbf{H}_{k}^H\mathbf{w}_k^{(t)}\right|^2+2\textup{Re}\left( ({\mathbf{w}_k^{H}}^{(t)}\mathbf{H}_{k}\mathbf{H}_{k}^H)(\mathbf{w}_k-\mathbf{w}_k^{(t)})\right)\\
&\triangleq\Pi_k, \forall k \in\mathcal{K}.
	\end{aligned}
\end{equation}

Hence, the subproblem for BS transmit beamforming optimization is approximated as the following problem
\begin{subequations}\label{problem formulation}
	\begin{align}	
		&\mathop{\rm{max}}\limits_{\mathbf{W}_{k},X_{k},Y_{k},y}\quad 
		2 \alpha \sqrt{y} -\alpha^{2} P_{tot}+\frac{\omega y}{P_{\max}}\\
		&\quad\,\;\;{\rm s.t.} \;\,\
		y\le\sum_{k\in\mathcal{K}}R_{k}^{lb},\\
		&\quad\,\qquad\quad
		R_{k}^{lb}\ge R_{\min}, \forall k \in\mathcal{K},\\
		&\quad\,\qquad\quad
		\frac{1}{X_{k}}\le\Pi_k, \forall k \in\mathcal{K},\\
		&\quad\,\qquad\quad
		\textup{(13b), (18), (21)},
	\end{align}
\end{subequations}
where $y$ is a slack variable. It can be found that, for fixed $\mathbf{ \Theta}_s$, problem (24) is a convex problem and therefore can be solved via existing optimization solvers, such as CVX~\cite{grant2014cvx}.
\vspace{-0.4cm}
\subsection{STAR-RIS Configuration} 
For given $\mathbf{w}_k$, problem (14) is reduced to finding the optimal STAR-RIS coefficient matrices $\mathbf{\Theta}_{s}$, which yields the following problem:
\begin{subequations}\label{problem formulation}
	\begin{align}
		&\mathop{\rm{max}}\limits_{\mathbf{\Theta}_{s}}\quad 
		2 \alpha \sqrt{\eta_{\rm SE}(\mathbf{\Theta}_{s})} 
		-\alpha^{2} P_{tot}(\mathbf{\Theta}_{s})+\frac{\omega \eta_{\rm SE}(\mathbf{\Theta}_{s})}{P_{\max}}\\
		&\;\;{\rm s.t.} \;\;\;\,\,
		\textup{(13c) -- (13e)}.
	\end{align}
\end{subequations}
Problem (25) is non-convex since it contains the non-convex constraints (13c) and (13d).

Define $\mathbf{Q}_k\triangleq{\textup{diag}}\left(\mathbf{h}_{k}^H\right)\in\mathbb{C}^{M \times M}$ and $\mathbf{D}_k\triangleq{\textup{diag}}\left(\mathbf{h}_{k}^H\right)\mathbf{G}\in\mathbb{C}^{M \times N}$, the SINR term in (4) can be written as Eq. (26). 
\begin{figure*}[h]
	\small
	\begin{equation}
		\begin{aligned}
			\textup{SINR}_{k}=\frac{\left|\mathbf{u}_{s_k}^H\mathbf{D}_k\mathbf{w}_{k}\right|^2}{\sum_{i\in\mathcal{K}\backslash \{k\}}\left|\mathbf{u}_{s_k}^H\mathbf{D}_k\mathbf{w}_{i}\right|^2+\sum_{l\in\mathcal{K}}\mathbf{u}_{s_k}^H\mathbf{D}_k\left(\kappa_u \mathbf{w}_l\mathbf{w}_l^H+\left(1+\kappa_u\right)\kappa_b {\widetilde{\textup{diag}}}\left(\mathbf{w}_l\mathbf{w}_l^H\right) \right)\mathbf{D}_k^H\mathbf{u}_{s_k}+(1+\kappa_u)\left(\left\|\mathbf{u}_{s_k}^H\mathbf{Q}_k \right\|^2\sigma_{A}^2+\sigma^2\right)}.
		\end{aligned}
	\end{equation}
	\vspace*{-28pt}
\end{figure*}
By lifting vectors $\mathbf{u}_s, \forall s\in\{t,r\}$ into matrix $\mathbf{U}_s={\mathbf{u}_s}{\mathbf{u}_s}^H\in\mathbb{C}^{M \times M}$, which satisfies the conditions:
\begin{gather}
			{\mathbf U}_s\succeq 0, \\
			\textup{Rank}({\mathbf U}_s)=1,\\
			{\widetilde{\textup{diag}}}({\mathbf U}_s)\le \rho_{\max}{\mathbf I}_{M},
\end{gather}
we can further express the SINR as Eq. (30), where $\mathbf{W}_k={\mathbf{w}_k}{\mathbf{w}_k}^H\in\mathbb{C}^{N \times N}$, $\widetilde{\mathbf{Q}}_k={\mathbf{Q}_k}{\mathbf{Q}_k}^H\in\mathbb{C}^{M \times M}$, and $\mathbf{E}_k=\sum_{l\in\mathcal{K}}\left(\kappa_u \mathbf{D}_k\mathbf{w}_l\mathbf{w}_l^H\mathbf{D}_k^H+\left(1+\kappa_u\right)\kappa_b\mathbf{D}_k {\widetilde{\textup{diag}}}\left(\mathbf{w}_l\mathbf{w}_l^H\right)\mathbf{D}_k^H \right)$.
\begin{figure*}[h]
	\normalsize
	\begin{equation}
		\begin{aligned}
			\textup{SINR}_{k}=\frac{\textup{Tr}\left(\mathbf{U}_{s_k}{\mathbf{D}_k} \mathbf{W}_k {\mathbf{D}_k^{H}}\right)}{\sum_{i\in\mathcal{K}\backslash \{k\}}\textup{Tr}\left(\mathbf{U}_{s_k}{\mathbf{D}_k} \mathbf{W}_i {\mathbf{D}_k^{H}}\right)+\textup{Tr}\left(\mathbf{E}_k\mathbf{U}_{s_k} \right)+(1+\kappa_u)\left({\textup{Tr}\left( \widetilde{\mathbf{Q}}_k\mathbf{U}_{s_k}\right)}\sigma_{A}^{2}+\sigma^2\right)},
		\end{aligned}
	\end{equation}
	\vspace*{-28pt}
\end{figure*}
Now, the SINR term is converted into a fractional function with a linear numerator over a linear denominator.
In the previous subsection, we have already demonstrated how to tackle the non-convex constraint (13d). Similarly, the rate expression here can be replaced with its lower bound as shown in (22). 

For the non-convex constraint (13c), the term on its left-hand side can be further rewritten as Eq. (31).
\begin{figure*}[h]
	\normalsize
	\begin{equation}
		P_{\rm act}=\sum_{s\in\{t,r\}}\textup{Tr}\left(\mathbf\Theta_{s}\left(\sum_{k\in\mathcal{K}}\mathbf G\mathbf{W}_k\mathbf{G}^H+\mathbf G\left(\kappa_b \sum_{k\in\mathcal{K}} {\widetilde{\textup{diag}}}\left(\mathbf{W}_k\right)\right)\mathbf G^H+\sigma_{A}^2\mathbf{I}_M\right)\mathbf\Theta_{s}^H\right).
	\end{equation}
	\hrulefill 
	\vspace*{-14pt}
\end{figure*}
With definition $\mathbf{\Upsilon}=\mathbf{I}_M\circ\left(\sum_{k\in\mathcal{K}}\mathbf G\left(\mathbf{W}_k+\kappa_b  {\widetilde{\textup{diag}}}\left(\mathbf{W}_k\right)\right)\mathbf{G}^H+\sigma_{A}^2\mathbf{I}_M\right)$, we have $P_{act}=\sum_{s\in\{t,r\}}\mathbf{u}_s^H \mathbf{\Upsilon} \mathbf{u}_s$. Then, the STAR-RIS amplification power constraint is equivalently transformed into the following convex constraint:
\begin{equation}
	\sum_{s\in\{t,r\}}\textup{Tr}\left(\mathbf \Upsilon\mathbf{U}_s \right)\le P_{\rm RIS}^{\max}.
\end{equation}

Based on the above reformulation, the subproblem for designing the STAR-RIS configuration can be rewritten as
\begin{subequations}\label{problem formulation}
	\begin{align}
		&\mathop{\rm{max}}\limits_{\mathbf{U}_{s},X_{k},Y_{k},y'}\quad 
		2 \alpha \sqrt{y'} -\alpha^{2} P_{tot}+\frac{\omega y}{P_{\max}}\\
		&{\rm s.t.}\;\;
		y'\le\sum_{k\in\mathcal{K}}R_{k}^{lb},\\
		&\;\;\;\quad R_{k}^{lb}\ge R_{\min}, \forall k \in\mathcal{K},\\
		&\;\;\;\quad\frac{1}{X_{k}}\le \textup{Tr}\left(\mathbf{U}_{s_k}{\mathbf{D}_k} \mathbf{W}_k {\mathbf{D}_k^{H}}\right), \forall k \in\mathcal{K},\\
		&\;\;\;\quad Y_{k}\ge \sum_{i\in\mathcal{K}\backslash \{k\}}\textup{Tr}\left(\mathbf{U}_{s_k}{\mathbf{D}_k} \mathbf{W}_i {\mathbf{D}_k^{H}}\right)+\textup{Tr}\left(\mathbf{E}_k\mathbf{U}_{s_k} \right) \notag\\
		&\;\;\;\quad+(1+\kappa_u)\left({\textup{Tr}\left( \widetilde{\mathbf{Q}}_k\mathbf{U}_{s_k}\right)}\sigma_{A}^{2}+\sigma^2\right), \forall k \in\mathcal{K},\\
		&\;\;\;\quad\textup{(27) -- (29), (32)}.
	\end{align}
\end{subequations}
The non-convexity of problem (33) is due to the rank-one constraint, i.e., $\textup{Rank}({\mathbf U}_s)=1, \forall s\in\{t,r\}$. By removing this constraint, problem (33) simplifies into a convex problem that can be efficiently solved using optimization solvers like CVX~\cite{grant2014cvx}. Moreover, to enforce the rank-one property of ${\mathbf U}_{t/r}$, the difference-of-convex (DC) relaxation method proposed in our earlier work~\cite{huang2022coexisting} can be employed. For conciseness, the detailed explanation is omitted here.
\vspace{-0.4cm}
\subsection{Overall Algorithm and Complexity Analysis}
\begin{algorithm}[t]\label{method1}
	\caption{Proposed Algorithm for Solving Problem (13)}
	\begin{algorithmic}[1]
		\STATE {\bf Initialize} $\mathbf{w}_{k}^{(0)}$, $\mathbf{\Theta}_{s}^{(0)}$, $\alpha^{(0)}$, and set $n=1$, threshold $\epsilon$.
		\STATE {\bf Repeat}
		\STATE \quad Set  $\mathbf{w}_{k}^{(n-1,t)}=\mathbf{w}_{k}^{(n-1)}$, $\mathbf{\Theta}_{s}^{(n-1,t)}=\mathbf{\Theta}_{s}^{(n-1)}$, and $t=0$.
		\STATE \quad {\bf Repeat}
		\STATE \quad\quad Obtain $\mathbf{w}_k^{(n-1,t+1)}$ by solving problem (24).\\
		\STATE \quad\quad Obtain $\mathbf{\Theta}_s^{(n-1,t+1)}$ by solving problem (33).\\
		\STATE \quad\quad $t\leftarrow t+1$.
		\STATE \quad {\bf Until} the objective function converges.
		\STATE \quad $\mathbf{w}_{k}^{(n)}\leftarrow \mathbf{w}_{k}^{(n-1,t)}$, $\mathbf{\Theta}_{s}^{(n)}\leftarrow \mathbf{\Theta}_{s}^{(n-1,t)}$.
		\STATE \quad Update $\alpha$ by (15).	
		\STATE \quad $n\leftarrow n+1$.
		\STATE {\bf Until} $\left|\alpha^{(n)}-\alpha^{(n-1)}\right|\le\epsilon$.
		\STATE {\bf Output} $\mathbf{w}_{k}^{\star}$, $\mathbf{\Theta}_{s}^{\star}$, and $\alpha^{\star}$.
	\end{algorithmic}
\end{algorithm}
\textbf{Algorithm 1} describes the proposed iterative method for solving problem (13) based on the quadratic transformation. In the outer layer, the update of $\alpha$ follows (15). While in the inner layer, an AO-based algorithm is developed to iteratively optimize the BS transmit beamforming and active STAR-RIS configuration. The computational complexity of \textbf{Algorithm 1} is summarized as follows: According to~\cite{luo2010semidefinite}, the complexity for designing the BS precoding in step 5 is $\mathcal{O}\left(\max \left(N, 3K+3\right)^{4} \sqrt{N} \log _{2}(1 / \varepsilon)\right)$, where $\varepsilon$ denotes the accuracy. The computational complexity for STAR-RIS configuration in step 6 with the interior-point method is $\mathcal{O}\left(\left(2M\right)^{3.5} \right)$. Therefore, the overall computational complexity of \textbf{Algorithm 1} is ${{\mathcal{O}}}\left(I_{\rm out}I_{\rm inn}\left( \max \left(N, 3K+3\right)^{4} \sqrt{N} \log _{2}(1 / \varepsilon)+\left(2M\right)^{3.5}\right)\right)$, where $I_{\rm inn}$ and $I_{\rm out}$ denote the number of iterations required to reach convergence in the inner and outer layers of \textbf{Algorithm 1}, respectively.
\section{Numerical Results}
This section presents numerical results to assess the effectiveness of the proposed scheme. The BS and the active STAR-RIS are fixed 40 meters apart in the simulations. Four users are randomly distributed within a circular region centered at the active STAR-RIS with a radius of $3$ meters. Specifically, there are two users on each side of the STAR-RIS. The active STAR-RIS has a total of $M=5\times6$ elements. All channels are assumed to adhere to the Rician channel model, with the relevant channel parameters referenced from~\cite{mu2021simultaneously}. The HWI coefficients at both transmitter and receivers are considered identical, i.e., $\kappa_b=\kappa_u=\kappa$. Simulation parameters are set to be $N=4$, $B = 10$ MHz, $P_{\rm RIS}^{\rm max} = 30$ dBm, $R_{\rm min}=0.4$ bit/s/Hz, $\xi=0.8$, $P_{r}=10$ dBm~\cite{huang2019reconfigurable}, $P_{c}=30$ dBm~\cite{liu2020energy}, $\sigma^2=\sigma_A^2 = -110$ dBm, $\kappa=0.02$, and $\rho_{\max}=5$.

\begin{figure}[t]
	\centering
	\includegraphics[width=3in]{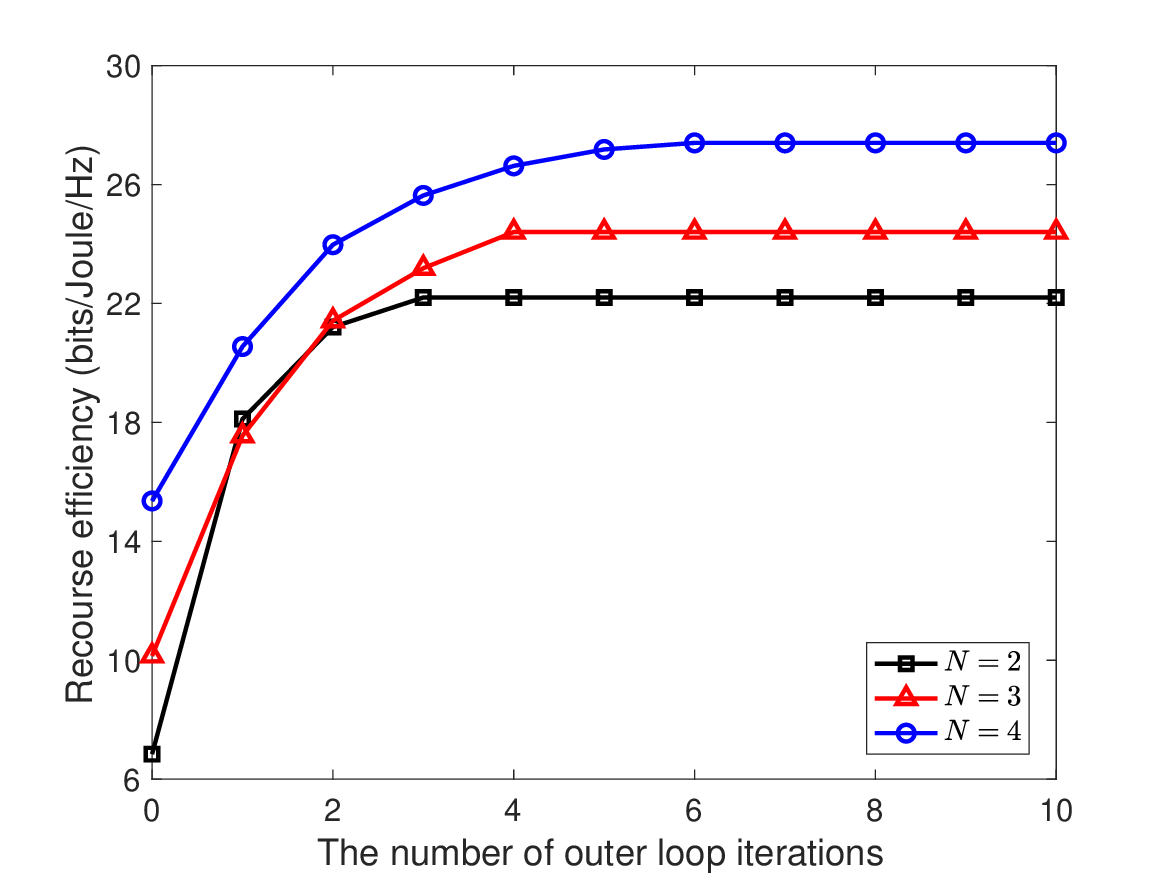}
	\vspace{-0.1cm}
	\caption{Convergence behavior of \textbf{Algorithm 1} under different numbers of BS antennas $N$.}
	\label{}
\end{figure}
\textbf{\textit{1) Convergence behavior of  \textbf{Algorithm 1}}:}
Fig. 2 depicts the convergence of \textbf{Algorithm 1} under different numbers of BS antennas $N$. As shown, the proposed algorithm converges rapidly, which highlights its effectiveness in terms of convergence performance. Moreover, we observe that while increasing $N$ may result in slower convergence due to the higher dimensionality and complexity of the optimization problem, it also significantly improves system RE. This improvement stems from the greater spatial degrees of freedom provided by additional antennas, allowing for more efficient resource allocation and ultimately boosting overall system performance.

\begin{figure}[t]
	\centering
	\includegraphics[width=3in]{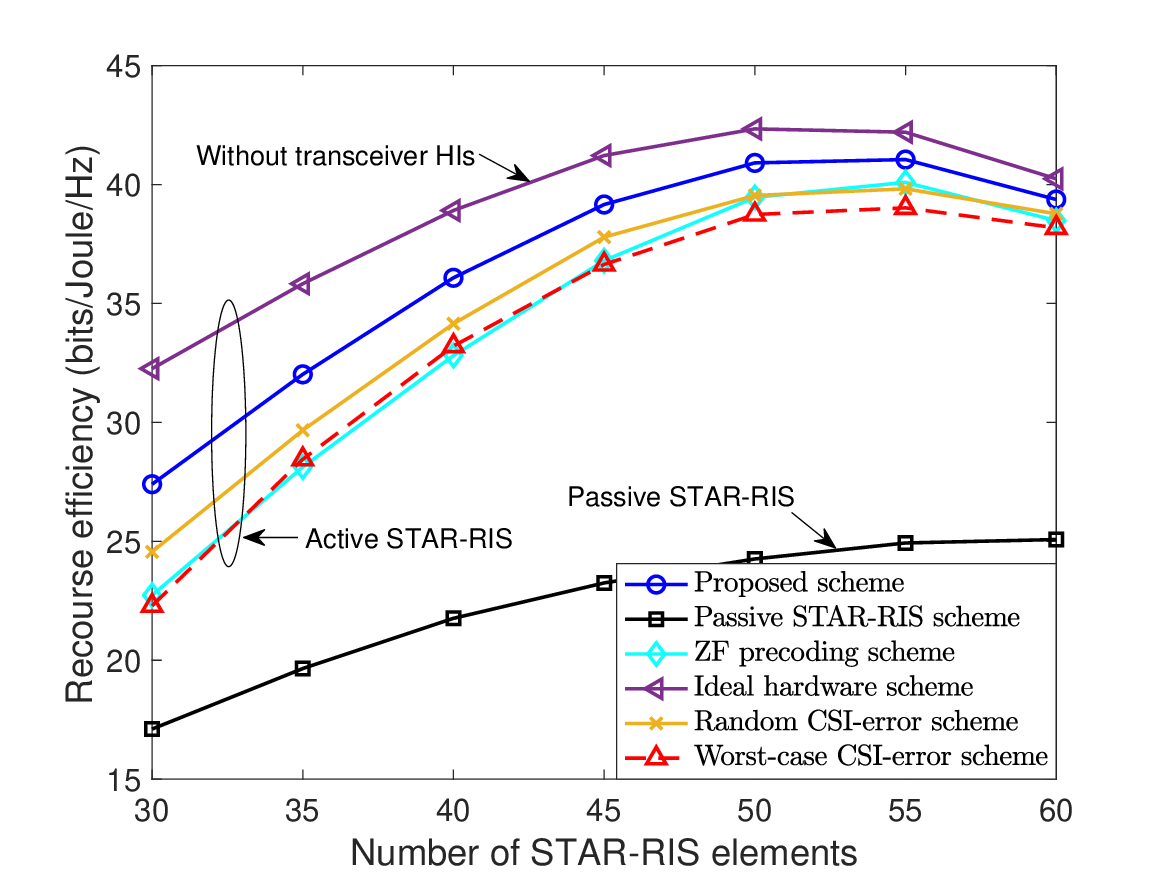}
	\vspace{-0.1cm}
	\caption{System RE versus the number of the STAR-RIS elements $M$.}
	\label{}
\end{figure}
\textbf{\textit{2) System RE versus STAR-RIS element number:}}
In Fig. 3, we explore the impact of the number of elements $M$ on the STAR-RIS when $\omega/P_{\max}=1$ and $P_{\rm BS}^{\max} = 30$ dBm. Here, four baseline schemes are considered: (1) \textbf{Passive STAR-RIS scheme}: In this case, we deploy a passive STAR-RIS in the network to assist signal transmissions; (2) \textbf{Zero-forcing (ZF) precoding scheme}: In this case, ZF precoding is adopted at the BS for transmit beamforming design; (3) \textbf{Ideal hardware scheme}:  In this case, the resource allocation algorithm is designed under the ideal hardware assumption, i.e., the HWI coefficient is set to $\kappa=0$; (4) \textbf{Random CSI-error scheme}: In this case, the system performance is evaluated under imperfect CSI, where the perturbed channels are modeled as $\widehat{\mathbf{G}} =\mathbf{G}+\Delta \mathbf{G}$ and $\widehat{\mathbf{h}}_{k} ={\mathbf{h}}_{k}+\Delta {\mathbf{h}}_{k}$. The errors $\Delta \mathbf{G}$ and $\Delta {\mathbf{h}}_{k}$ satisfy $\|\Delta \mathbf{G}\|_{F}\le\rho_c\|\mathbf{G}\|_{F}$ and $\|\Delta{\mathbf{h}}_{k}\|_{F}\le\rho_c\|{\mathbf{h}}_{k}\|_{F}$, respectively, where $\rho_c$ is randomly drawn from $[0, 10^{-2}]$ in each channel realization~\cite{wu2021channel}; and (5) \textbf{Worst-case CSI-error scheme}:  This scheme reflects the most adverse case of imperfect CSI within the bounded-error model. The perturbed channels are constructed such that $\|\Delta \mathbf{G}\|_{F} = 10^{-2}\|\mathbf{G}\|_{F}$ and $\|\Delta {\mathbf{h}}_{k}\|_{F} = 10^{-2}\|{\mathbf{h}}_{k}\|_{F}$.

It can be observed that the best system performance is obtained when transceiver HWIs are ignored. Compared to our proposed scheme, which optimally balances signal gain and system energy consumption, the ZF precoding scheme shows suboptimal performance. The reason is that ZF focuses solely on interference cancellation, leading to less efficient resource utilization, especially in multiuser scenarios. Furthermore, we found that in the active STAR-RIS schemes, the system RE exhibits an initial increase with $M$, followed by a decrease. This behavior contrasts with the results in schemes relying on passive STAR-RISs, where a larger $M$ consistently leads to higher RE. When $M$ is relatively small, benefiting from the array gain introduced by the STAR-RIS, appropriately expanding its scale helps achieve desirable performance. However, if the STAR-RIS is active, a large number of elements requires higher power demand, potentially leading to a degradation in system performance. As demonstrated by our results,  imperfect CSI further exacerbates the performance deterioration. In particular, both the random and worst-case CSI-error schemes result in a reduced RE compared to the perfect CSI case, with the latter yielding the most severe performance loss due to the high CSI error. Future work will therefore focus on developing robust optimization algorithms against the CSI uncertainty.

\begin{figure}[t]
	\centering
	\includegraphics[width=3in]{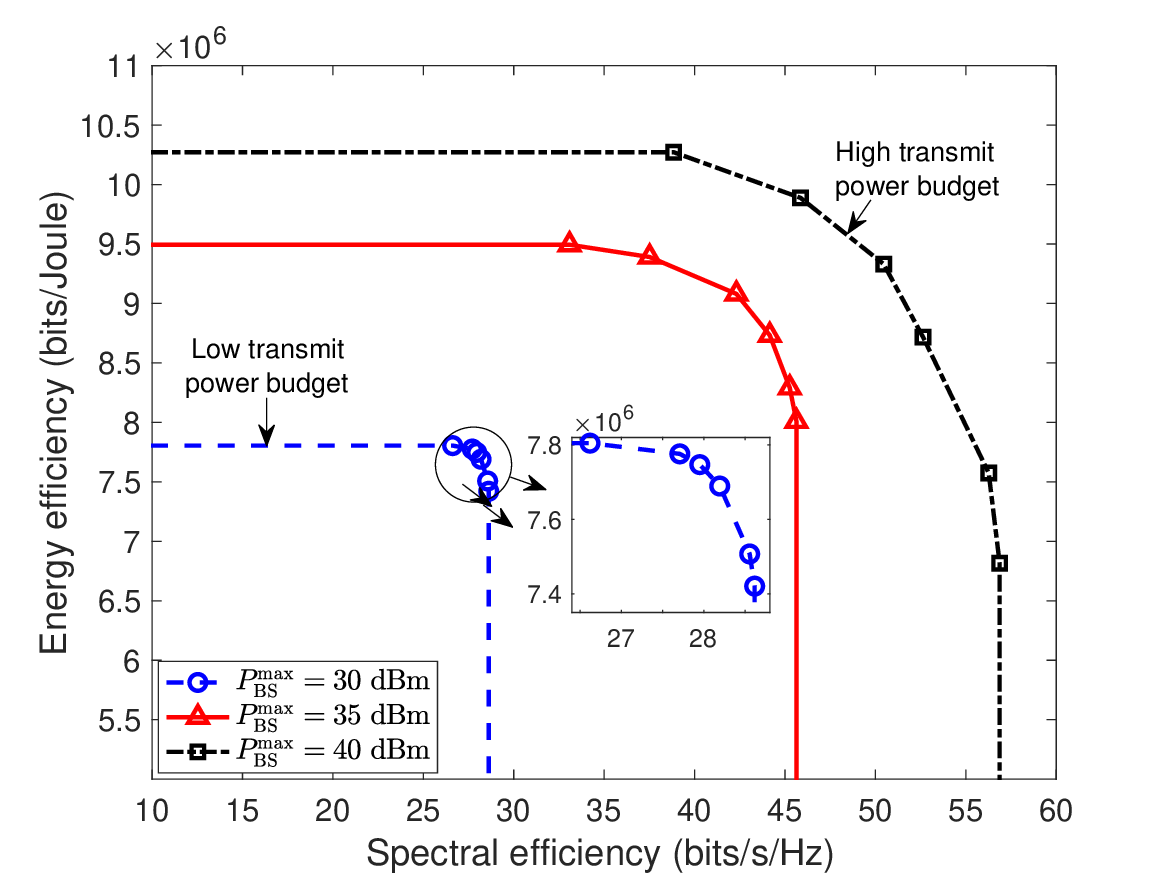}
	\vspace{-0.1cm}
	\caption{SE-EE tradeoffs under different transmit power budgets $P_{\rm BS}^{\max}$.}
	\label{}
\end{figure}
\textbf{\textit{3) System SE-EE tradeoff:}}
Fig. 4 characterizes the tradeoff between system SE and EE under various transmit power budgets $P_{\rm BS}^{\max}$. The results are obtained by varying the weighting factor $\omega$ across different random channel realizations. As shown in this figure, increasing $P_{\rm BS}^{\max}$ helps to achieve a larger SE-EE region. Specifically, when $P_{\rm BS}^{\max}$ is low, the system strives to utilize all available power resources to enhance performance, regardless of how the weighting factor is adjusted, the performance in terms of SE and EE remains nearly identical. By contrast, when $P_{\rm BS}^{\max}$ is high, depending on the value of $\omega$, the proposed algorithm for maximizing system RE presents diverse tradeoffs between SE and EE.
\section{Conclusions}
In this work, we investigated an active STAR-RIS assisted multiuser communication system considering the presence of transceiver HWIs. A RE maximization problem was formulated to jointly optimize the BS precoding and active STAR-RIS beamforming. Numerical results demonstrated the effectiveness of the proposed framework in enhancing system performance. Moreover, a flexible tradeoff between system SE and EE was revealed. By appropriately managing their nonlinear relationship, significant improvements in overall system RE can be achieved.
\bibliographystyle{IEEEtran}
\bibliography{mybib}
 \end{document}